# Asymmetric Micro-Doppler Frequency Comb Generation via Magneto-Electric Coupling

Dmitry Filonov[1,2], Ben Z. Steinberg[1] and Pavel Ginzburg[1,2,*]

[1]School of Electrical Engineering, Tel Aviv University, Tel Aviv, 69978, Israel
[2]ITMO University, St. Petersburg 197101, Russia

**Abstract**:

Electromagnetic scattering from moving bodies, being inherently time-dependent phenomenon, gives rise to a generation of new frequencies, which could characterize the motion. While a standard linear path leads to a constant Doppler shift, accelerating scatterers could generate a micro-Doppler frequency comb. Here, a spectra produced by rotating objects, was studied and observed in a bi-static 'lock in' detection scheme. Internal geometry of a scatterer was shown to determine the spectra, while the degree of structural asymmetry was suggested to be identified via signatures in the micro-Doppler comb. In particular, hybrid magneto-electric particles, showing an ultimate degree of asymmetry in forward and backward scattering directions were investigated. It was shown that the comb in the backward direction has signatures at the fundamental rotation frequency and its odd harmonics, while in the forward scattered field has the prevailing peak at the doubled frequency and its multiples. Additional features in the comb were shown to be affected by the dimensions of the particle and strength of magneto-electric coupling. Experimental verification was performed with a printed circuit board antenna, based on a wire and a split ring, while the structure was illuminated with at 2GHz carrier frequency. Detailed analysis of micro-Doppler combs enables remote detection of asymmetric features of distant objects and could find use in a span of applications, including stellar radiometry and radio identification.

*pginzburg@post.tau.ac.il

## Introduction

Investigation of electromagnetic processes in moving coordinate systems requires applying a certain set of transformations to laws of electrodynamics, formulated for reference frames at rest [1]. Relativistic effects could substantially change the regular form of Maxwell's equations, material constitutive relations and boundary conditions, especially in the case of accelerated motion, which, in the most general case, requires applying the formalism of the general theory of relativity [2]. Nevertheless, the majority of practical applications take place in regimes of slow motion and acceleration, where simplifying approximations could be applied. One of the practical examples, dealing with non-relativistic scenarios, is radio detection and ranging (Radar), where distant objects are probed with short pulses. Time delay between transmitted and received pulses enables extracting the range, while Doppler shifts hold information about velocities [3]. Approximate electromagnetic analysis of the phenomenon relies on a set of static simulations, stitched together along a mechanical path of an object – this approach will be referred here as *adiabatic*. Full characterization of the mechanical motion also requires knowledge of accelerations, which could be identified with additional spectral analysis. For example, techniques for detection of helicopter propellers (axial rotation implies accelerated motion) do exist [4],[5],[6]. Accelerating bodies, in contrast to uniformly moving objects, produce much richer spectral signatures, called micro-Doppler shifts [7]. Recently, micro-Doppler combs, generated by axially rotating wires and split ring resonators (SRRs), were investigated [8],[9]. In particular, objects illuminated by a field of frequency ω, generate a frequency comb at a scattered field - the peaks are equidistant and situated at $\omega \pm n\Omega$, with $\Omega$ being the angular frequency of rotation and $n$ is an integer number. It is worth underlining, that those additional spectral features are generated by time-varying boundary conditions. Furthermore, it was shown that phase retardation effects along a rotating scatterer control amplitudes of the peaks in the comb. Along with the phase accumulation effects, the symmetry properties of an object have crucial importance. For example, scatterers with reflection symmetry and rotating around their centers cannot generate even peaks in the comb. Degree of asymmetry is important for remote characterization of distant moving objects and its impact will be investigated here on an example of an asymmetric scatterer.

Here a frequency comb, produced by asymmetric rotating objects, is studied theoretically, numerically and observed experimentally with a 'lock in' detection scheme. It is shown, that micro-Doppler combs, generated by asymmetric scatterers, strongly depend on the observation direction. In particular, micro-Doppler frequency comb generation from a hybrid magneto-electric particle (HMEP), consisting of a split ring resonator (SRR) and a thin wire is studied. The HMEP has asymmetric backscattering when illuminated by a plane wave from opposite directions [10] and, as the result, it shows completely different micro-Doppler combs, detected at forward and backward directions (Fig. 1).

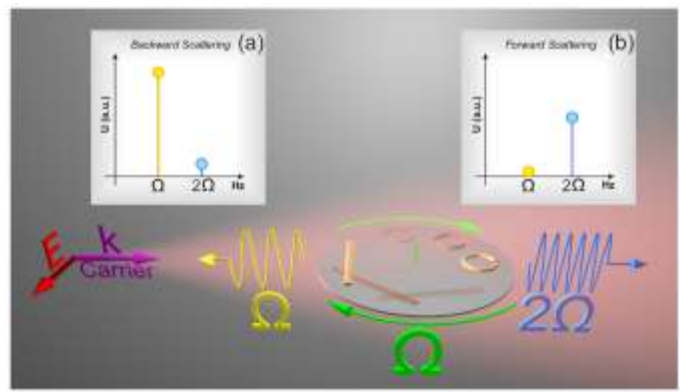

**Fig. 1.** Illustration of an asymmetric micro-Doppler comb generation from axially rotating magneto-electric particle (HMEP). Split ring resonator and a wire, forming the HMEP, are rotated with an angular frequency Ω in the polarization plane of the incident wave. Forward and backward scattered fields demonstrate different micro-Doppler spectra (insets illustrate frequencies at the baseband).

## Theoretical Formulation

Scattering from objects of deep subwavelength dimensions could be described by a *single* leading dipolar terms in the multipolar decomposition [11]. In this case, it could be shown that the field scattered from this object, being rotated around it's center (position of the effective point dipole), will have only one additional harmonic at $\pm 2\Omega$ ($\Omega$ is the angular rotation frequency) and it does not depend on the position of the detector (e.g. [12], [8]). Objects with nontrivial internal structure could generate an entire frequency comb of micro-Doppler shifts with relative amplitudes, dependent on the detection direction. In order to demonstrate this effect, a scatterer, which electromagnetic analysis requires at least two leading dipole terms in the multipolar decomposition will be considered. While two dissimilar electric dipoles, separated by a wavelength-comparable distance could possess asymmetric back scattering, being illuminated from opposite sides, combination of electric and magnetic dipoles was recently shown to provide an ultimate degree of asymmetry[10]. The hybrid magneto-electric particle (HMEP) consists of a SRR and a thin wire (Fig. 1) - electromagnetic response of the first one could be approximated with a magnetic dipole (MD), while the second constitutive element acts as an electric dipole (ED). The asymmetric reflection of the HMEP stems from the retarded coupling mechanism between ED and MD, depends on illumination wave frequency, inter-particle distance, and the individual parameters of the ED and MD. As the result, backward scattering strongly depends on the particle's orientation with respect to the incident wave k-vector. Consequently, the signal has the periodicity of $2\pi$ (full angle of rotation), if $\Omega$ is normalized to 1. On the other hand, the scattered field in the forward direction has $\pi$-periodicity, as the result of the Lorentz reciprocity principle applied on particles that obey time-reversal symmetry. Therefore, $\theta$ and $\theta+\pi$ angles (HMEP major axis in respect to the k-vector) provide the same forward scattered signal. As the result, micro-Doppler comb in forward and backward directions will have completely different amplitudes of the peaks. It is worth noting in brief, that similar behavior is expected from other particles too. For example, the so-called Huygen's elements, relying on interference between electric and magnetic dipolar responses that suppress the backward scattering [13],[14], are expected to provide a similar behavior. The same applies on meta-particles with nonsymmetrical scattering (e.g. omega, omega-Tellegen and the chiral-moving structures), which are discussed in details in [15],[16],[17]. It is worth emphasizing, however, that in these additional examples the electromagnetic description of the particle at rest still requires both, electric and magnetic dipoles.

*Adiabatic theory*

In order to analyze the micro-Doppler comb, obtained in forward and backward directions, adiabatic approximation will be employed. Here the set of static configurations will be analyzed and the results will be stitched together, providing a $2\pi$-periodic function. This function (either forward or backward scattering) will be decomposed into Fourier series, where the basic (smallest) discrete frequency corresponds to the angular velocity of the rotation.

The detailed analytical formulation of scattering from two discrete dipoles, forming the HMEP, was reported at [10]. While the case of 0 and $\pi$ orientation of the particle with respect to the incident wave direction the scattered fields have short analytical expression, the case of an arbitrary angle is described by cumbersome expressions, resulted from matrices inversions. In the point dipole description, attributed to the geometry mentioned above, the polarizabilities of the particles are given by:

$$\ddot{\alpha}_e = \alpha_{Wire} \begin{pmatrix} \cos^2(\theta) & \frac{\sin(2\theta)}{2} & 0 \\ \frac{\sin(2\theta)}{2} & \sin^2(\theta) & 0 \\ 0 & 0 & 0 \end{pmatrix}; \quad \ddot{\alpha}_m = \alpha_{SRR} \begin{pmatrix} 0 & 0 & 0 \\ 0 & 0 & 0 \\ 0 & 0 & 1 \end{pmatrix}, \quad (1)$$

where $\ddot{\alpha}_e$ and $\ddot{\alpha}_m$ are polarizability tensors of ED and MD, correspondingly, $\alpha_{Wire}$ and $\alpha_{SRR}$ are intrinsic coefficients of the wire and SRR at rest (solely defined by their geometry), while $\theta$ is the tilting angle of the HMEP in respect to the incident wave k-vector. The structure lays the X-Y plane and the rotation vector is parallel to Z (Fig. 2(c)). MD tensor is assumed to be independent on the rotation angle, which is a reasonable approximation even for a single-slit SRR [18] (further symmetrization could be done with the double-slit geometry, e.g. [19]). On the other hand, ED tensor has two components, as the projection of electrical field on the wire does depend on the orientation (pay attention that the polarizability of the wires cannot change the sign with an angle). In principle, a HMEP made of a pair of isotropic ED and MD, will show a similar micro-Doppler behavior to the one discussed hereafter. Adiabatic theory is based on the matrix formulation [10] and polarizability tensors, taken from Eq. 1. The results are summarized on Fig. 2. The dimensions of actual elements are (i) ring radius is $R_{SRR}$=9.5mm, slit width is 1 mm (ii) wire length is $L_{Wire}$=60mm and used in all of the subsequent studies. All the elements were modeled as printed board elements on FR4 substrates. Polarizabilities $\alpha_{Wire}$ and $\alpha_{SRR}$, extracted from these parameters appear in Fig. 3(a, b). Fig. 2 summarizes typical results for HMEP with a separation distance d=38mm between the wire and the ring. The structure is illuminated with the 1.96GHz carrier wave. Those parameters correspond to the maximal static asymmetry factor, as will be discussed later. Fig. 2(a, b) shows the micro-Doppler comb for the given particle. While the first odd peak at $\Omega$ is the most significant one in the backwards scatted field (panel a), it completely vanishes in the forward direction (panel b). This behavior is quite fundamental and follows from the Lorenz reciprocity theorem, which predicts the signal to be an even function in $\Omega$ ($\theta$ and $\theta+\pi$ angles of the particle in respect to the k-vector lead to the same scattering). Consequently, all even peaks in the forward comb vanish, as it can be seen from Fig. 2(b). $\Omega$ and $2\Omega$ peaks in the backward direction are of the main importance in this work (Fig. 2(a)). This behavior could be directly recalled by observing the polar plots, showing absolute values of far field amplitude in both forward and backward directions (the angle corresponds to the orientation of the particle's symmetry axis in respect to the k-vector of the incident wave) (Fig. 2(d)). While the forward scattering diagram poses complete symmetry (in respect to the vertical axis on the diagram), the backward case is completely asymmetric, showing the absolute values of scattering amplitudes in a range from a maximum to an almost complete zero. Side lobes (in general, deviations from cosine profile) are responsible for creation of higher order peaks, appearing on panels (a) and (b).

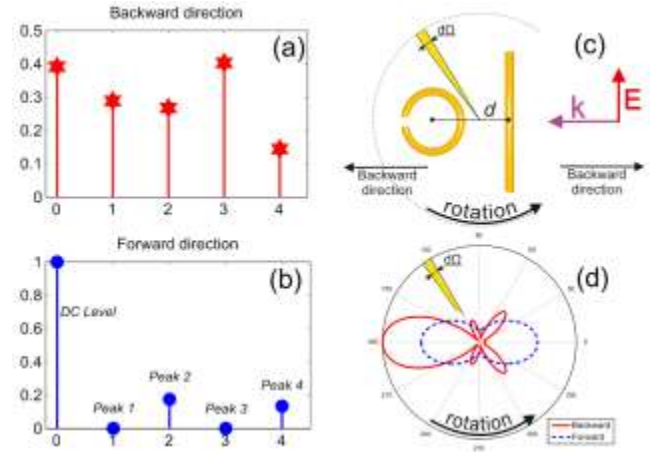

**Fig. 2.** Nonlinear frequencies generation by axially rotating magneto-electric particle (HMEP) – analytic theory. Micro-Doppler frequency comb in backward (a) and forward (b) directions. Particle parameters are d = 38 mm, $R_{SRR}$ = 9.5mm, $L_{Wire}$ = 60mm, the illumination frequency is 1.96 GHz. (c) Schematics of the HMEP particle and the illumination field layout. (d) Polar plots of absolute values of scattered far field amplitudes, as the function of the particle's orientation with respect to the incident wave k-vector. Red-solid line – backward direction; blue-dashed line – forward.

Fig. 2 shows the characteristic behavior of the micro-Doppler comb for one particular realization. In order to investigate the impact of particle's geometry on the comb, scattering properties of HMEP in the static regime are to be recalled (this analysis follows the approach reported at [10]). First, polarizabilities of constitutive elements are retrieved (Fig. 3(a, b)) and used as parameters in the subsequent investigations. As it was previously discussed, the unbalance in backscattering from the particle, oriented at 0 and π angle, is responsible for creating odd peaks. Static asymmetry factor (visibility) summarizing this behavior, is defined as:

$$V_a = \frac{\left|E_{sc}^B(\theta=0)\right|^2 - \left|E_{sc}^B(\theta=\pi)\right|^2}{\left|E_{sc}^B(\theta=0)\right|^2 + \left|E_{sc}^B(\theta=\pi)\right|^2}, \quad (2)$$

where $E_{sc}^B$ are the far-fields of back-scattered signals, obtained for opposite orientation of the HMEP. Color plot of $V_a$, as the function of separation distance between the elements and the illumination frequency appears on Fig. 3(c). It can be clearly seen, that the static asymmetry factor has clear resonance structure, corresponding to both resonances of constitutive elements and the separation distance between them. For example, the parameters used for obtaining results on Fig. 2, correspond to one of the maxima of the static asymmetry factor (bounded between -1 and 1, by the definition (Eq. 2)).

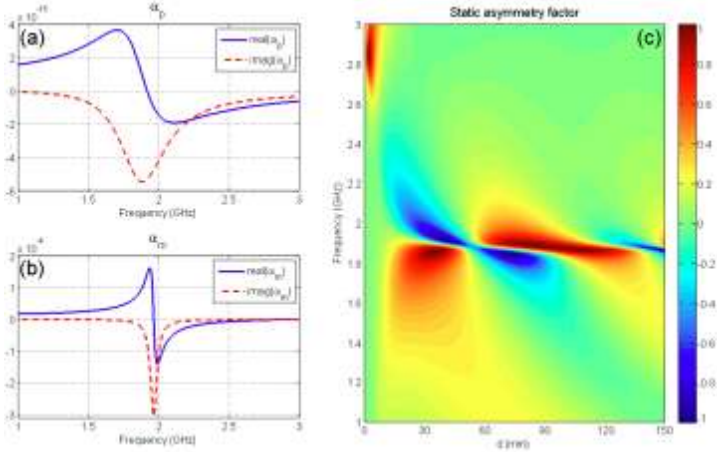

**Fig. 3.** Properties of the static configuration. (a) polarizability of the wire, as the function of frequency. (b) polarizability of the split ring, as the function of frequency. (c) color map of the static asymmetry factor, as the function of frequency (y-axis) and separation distance between the wire and the ring (x-axis).

Adiabatic approach for deriving micro-Doppler combs enables investigating properties of the peaks as the function of system's parameters. First four peaks of the comb in the backscattered signal appear on Fig. 4. Color maps, normalized to the same value, are represented as the function of the same parameters, used for representation of Fig. 3. Fig. 2(a) provides one point on each color map of Fig. 4. First, it can be seen that the micro-Doppler peaks are stronger at the regions where HMEPs poses strong static asymmetry. This behavior is rather expected, as the structure scatters stronger at those sets of parameters. The first peak is much broader in the parametric space than all the rest, as it encapsulates major impact of the imbalance between scattering from HMEP, oriented at 0 and π angles.

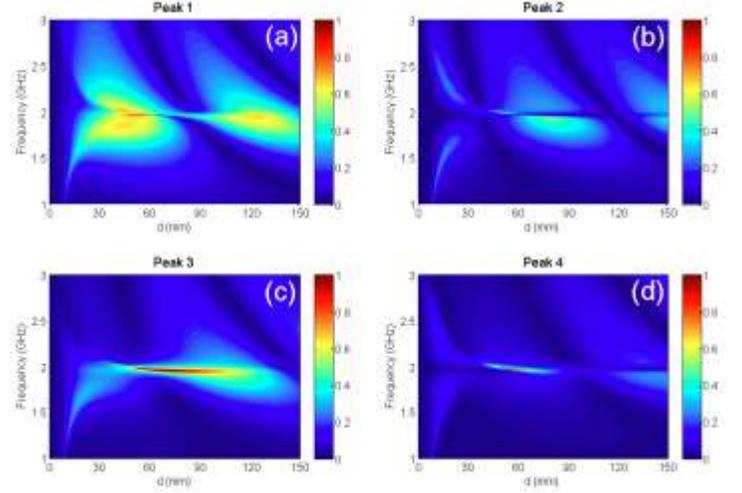

**Fig. 4.** Amplitudes of the micro-Doppler peaks in the backscattered signal from hybrid magneto-electric particles, as the function of illumination frequency (y-axis) and separation distance between the constitutive elements (x-axis). (a) first, (b) second, (c) third, (d) forth peaks in the comb. All the color maps are normalized to the same value.

Higher order peaks are capable of fast-oscillating features on the polar scattering diagrams (e.g. Fig. 2(d)) and, consequently, are more influenced by resonant effects. As the result, higher order peak in the comb (Fig. 4(b-d)) are narrower. Finally, higher order peaks are sensitive to the retardation effects (recall, that a point dipole has a single second order peak). It can be seen that amplitudes of higher order peaks (e.g. third and fourth) are more pronounced at larger separation distance (d) between the constitutive elements of HMEPs.

*Numerical Modeling*

While the compact analytical model enables investigating the influence of a wide range of geometrical parameters on the micro-Doppler comb, it does not take into account effects of substrates. The presence of those additional layers with complex dielectric constants affects both polarizabilities on HMEP constitutive elements and the coupling between them. Several semi-analytical models for addressing those effects do exist [10]. However, full wave numerical simulation enables taking into account all the relevant effects (including finite sizes of constitutive elements) on expense of the computational time. Numerical model, based on finite elements method, follows exactly the same approach as the analytical one – values of scattered complex amplitudes, as the function of the static particle at an angle to the incident wave direction, are recorded and then Fourier-transformed. After performing a set of numerical optimisations (straightforward and not shown here) the particles' dimensions were chosen to be d = 38 mm, $R_{SRR}$ = 9.5mm, $L_{Wire}$ = 60mm (the same values were chosen in the theory section for this exact reason). Material parameters for thin copper strips and RF4 substrate were taken from widely available sources.

Fig. 5 summarizes the results, shown for three different carrier frequencies – (i) 1.96GHz, where the particle has maximal backward-forward scattering asymmetry, (ii) 1.6GHz, and (iii) 2.6GHz, where the particle is off resonance and the forward-backward difference in micro-Doppler is expected to vanish. Insets to all the panels of Fig. 5 show the angular dependent plot, which is taken as the basis for the Fourier transform – the peaks at normalized frequencies appear on the main panels. It can be seen from the plots that the odd peaks are present only in the back scattering signal and only at the resonance of the HMEP particles. All other cases show signatures on even harmonics only, while the odd ones are either on the numerical noise level (in the case of the forward direction) or small compared to the main contributions. Similar behaviour was predicted by the analytical model. This very high sensitivity of the odd peaks to the resonant properties of the structure could serve as an additional channel for its remote probing.

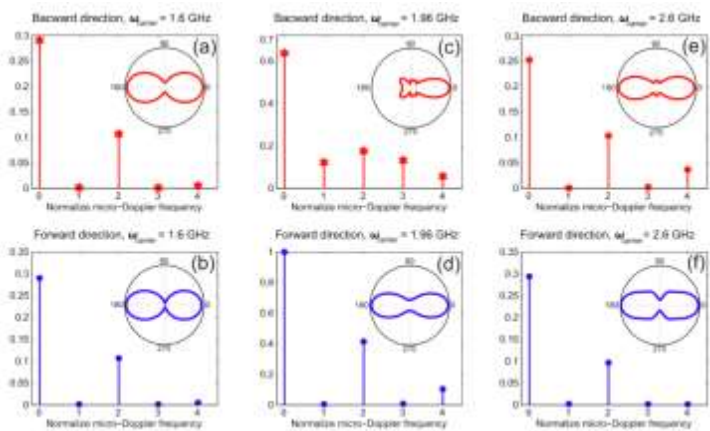

**Fig. 5.** Numerical modelling of micro-Doppler combs from rotating HMEP particle on FR4 substrate - amplitudes for Fourier peaks at normalized frequencies. Upper raw (a,c,e) – backward direction, lower raw (b,d,f) – forward direction. Columns of the figure correspond to different carrier frequencies – (a, b) 1.6GHz, (c,d) 1.96GHz, where the particle has the maximal backward-forward static scattering asymmetry, (e,f) 2.6GHz. Insets to all the panels are angular-dependent scattering (polar angle is between the symmetry axis of HMEP particle and the k-vector of the incident wave).

**Experimental Results**

Micro-Doppler combs in backward and forward directions are studied experimentally. The experiment was conducted in an anechoic chamber where the HMEP was positioned in the far fields of receiving and transmitting antennas. HMEP particle was fabricated by chemical etching of the thin copper layer, deposited on FR4 substrate (standard printed board circuit design – photograph of the sample is on the inset to Fig. 6(a)). HMEP was mounted on a motor's shaft, stabilized with a 3D-printed plastic rack. The motor is controlled by a DC current, enabling rotational speeds of up to hundreds of hertz. A constant angular frequency $\Omega = 2\pi \times 14\ Hz$ was chosen for the experiment. The whole structure was placed between two horn antennas, separated by a 2m distance, ensuring that both the transmitting and receiving horns are at the far field region of the scatterer. A Vector Network Analyzer (VNA) (Agilent E8362B), was used as a source of a 1.96GHz continuous wave, feeding the transmitting antenna with the polarization set in the Y-direction. The output of the receiver (either a horn antenna in front or the same antenna for collecting the backward scattering) was amplified and down-converted to the baseband by mixing it with the same 1.96GHz carrier, as in the incident field. The output of the mixer was passed through a low pass filter to remove residual high frequencies and directed to a lock-in amplifier where the low-frequency spectral shifts of the scattered field were recorded. The backscattering was extracted from the measured complex-valued signals, obtained in several steps in order to eliminate instrumental responses. The calibration was performed with a large area metallic mirror, having a reflection coefficient equal to -1. All the reflected signals were normalized accordingly to factorize the impact of the measurement apparatus. In the case of the dynamical experiments residual micro-Doppler shifts of a motor without the sample were subtracted from the data in order to factor out influences of mechanical parts.

Fig. 6 summarizes the experimental results. Panels (a) and (b) show micro-Doppler combs at backward and forward directions. It can be seen that the predominating peak in the forward case is indeed even, while the entire comb is observed in the backward direction with the full correspondence with the theoretical predictions. Residual odd peaks on panel (b) originate from mechanical instabilities, e.g. small drift in the rotation speed and precession of the motor shaft. Panels (c and d) show experimental results for a simple thin rotating wire (photograph is on the inset), mounted on the same system. This investigation serves as a calibration test. Symmetrical structure, such as a wire, is not expected to demonstrate odd harmonics in either forward or backward directions. The experimental data confirms this prediction. It is worth noting, that back scattering experiments are much sensitive to noise. Signals collected in the forward direction have much better signal to noise ratios. Furthermore, micro-Doppler harmonics are sensitive to even slight displacement of the object from the symmetry. Higher order peaks are also influenced by the drifting angular velocity of the motor (instabilities) and parasitic precessions of the motor shaft. All those parameters have influences on the micro-Doppler combs, especially in the case of resonant systems, where noise could be amplified.

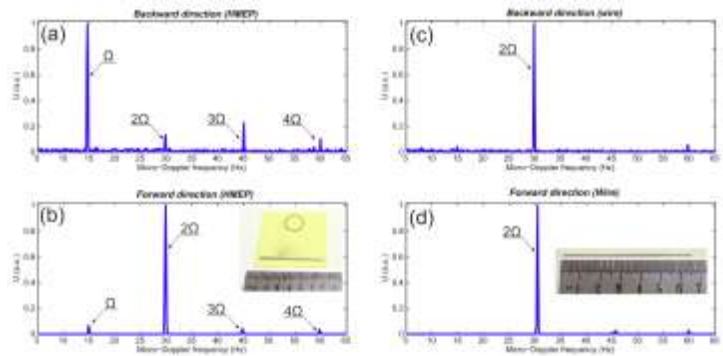

**Fig. 6.** Experimental observation of micro-Doppler combs. (a, b) – HMEP particle; (c,d) –thin metal wire. Upper raw – backward direction, lower raw – forward direction. Angular frequency of mechanical rotation is $\Omega = 2\pi \times 14\ Hz$. Insets – photographs of the samples.

**Outlook and Conclusions**

Electromagnetic scattering from axially rotating magneto-eclectic particles was studied analytically, numerically and experimentally. It was shown, that rotational (accelerated) motion gives rise to micro-Doppler frequency comb generation. Properties of those combs were studied by considering two detection directions (bi-static configuration). In particular, forward (collecting antenna is in front of the source) and backward (collection and illumination is performed with the same antenna) cases were investigated. It was shown that collection in backward direction provides more information about the scatterers. In particular, a collection of odd micro-Doppler harmonics could be missing in the forward case owing to Lorenz reciprocity principle, while the backward detection does not pose those limitations. Furthermore, resonant properties and retardation effects were shown to have signatures in higher-order micro-Doppler harmonics.

Characterization of rotating electromagnetic scatterers by means of micro-Doppler spectra in bi-static and multi-static configurations (several detection directions) could serve as a powerful tool in remote sensing and spectroscopy. Careful analysis, utilizing mechanical motion of detected objects, enables extracting extended information on their internal structure, similar to approaches used in synthetic aperture radars (SARs).

**Acknowledgments**

This work was supported, in part, by TAU Rector Grant, PAZY foundation, KAMIN project, and German-Israeli Foundation (GIF, grant number 2399). Numerical calculations were supported in part by the Russian Fund for Basic Research within the project 16-52-00112. The calculations of magnetic field distributions and multipole moments has been supported by the Russian Science Foundation Grant No. 16-12-10287.